\documentclass[preprint,showpacs,preprintnumbers,amsmath,amssymb,superscriptaddress,revsymb,prb]{revtex4}
\usepackage[dvips]{graphicx}
\usepackage{mathptmx}

\begin{document}
\title{Quantum conductance of homogeneous and inhomogeneous interacting 
electron systems}
\author{P. Bokes } \email{peter.bokes@stuba.sk}
\affiliation{Department of Physics, Faculty of Electrical Engineering and
        Information Technology, Slovak University of Technology, 
	Ilkovi\v{c}ova 3, 812 19 Bratislava, Slovak Republic}
\author{J. Jung}
\affiliation{Physics Division, National Center for Theoretical Sciences,  
	P.O. Box 2-131, Hsinchu, Taiwan}
\affiliation{Department of Physics, University of York, Heslington, York
         YO10 5DD, United Kingdom}
\author{R. W. Godby}
\affiliation{Department of Physics, University of York, Heslington, York
         YO10 5DD, United Kingdom}

\date{\today{}}

\begin{abstract}
We obtain the conductance of a system of electrons connected to leads, within time-dependent density-functional theory, using a direct relation between the conductance and the density response function. 
Corrections to the non-interacting conductance appear as a consequence of the functional form of the exchange-correlation kernel at small frequencies and wavevectors. 
The simple adiabatic local-density approximation and non-local density-terms in the kernel both give rise to significant corrections in general.  In the homogeneous electron gas, the former correction remains significant, and leads to a failure of linear-response theory for densities below a critical value.
\end{abstract}


\maketitle
Time-dependent density-functional theory extends the domain of {\it ab-initio} calculations to systems carrying a current, but relies on the accuracy of the exchange-correlation energy functional of the electron density (at present and past times), which in practice has to be approximated.
Impressive success has been achieved within the non-equilibrium Green's function formulation using the simple ground-state density-functional
exchange-correlation potential in a self-consistent 
formulation~\cite{Taylor02, Basch05} (gDFT).  However, limitations of the 
latter approximation were recently identified~\cite{Sai05,Toher05,Burke05,Palacios05,Stefanucci04,DiVentra04b}. For instance, gDFT's omission 
of the derivative-discontinuity in the exchange-correlation energy functional was 
found responsible for serious errors in transport calculation through localized resonant levels~\cite{Toher05,Burke05}. 
Improvements through an unrestricted gDFT formulation have been argued to describe properly some aspects of the Coulomb blockade in quantum junctions~\cite{Palacios05}.
At this level of the theory the exchange-correlation potential of the equilibrium system, $v_{xc}$, is responsible for the electron interaction effects.

In a further theoretical development, Na Sai {\it et al.}~\cite{Sai05} identified a dynamical correction 
to the resistance of a quantum junction stemming from the contribution of the exchange-correlation electric field to the overall drop in the total potential, as reflected in
the exchange-correlation kernel $f_{xc}$. They estimated the correction within time-dependent 
current-density functional theory~\cite{Vignale97} (TDCDFT) and showed that 
it has its origin in the non-local density-dependence of the functional. 
The very applicability 
of time-dependent density-functional theory (TDDFT) to the problem 
of quantum transport in the long-time limit has been discussed 
in depth by G. Stefanucci and C.-O. Almbladh~\cite{Stefanucci04} 
and by M. Di Ventra and T. Todorov~\cite{DiVentra04b}. 

Several 
authors have proposed alternative treatments that avoid the complexities 
of the exchange-correlation kernels of TD(C)DFT, either by using  
the usual gDFT approach in combination with a model 
self-energy within the central region~\cite{Ferretti05}, or by treating 
the central region with the configuration integration method~\cite{Delaney04} 
while approximating the non-equilibrium distribution of the electrons.

In this paper we address the class of corrections to the linear-response conductance that arise
from the exchange-correlation
kernel and potential in TDDFT.  A kernel must give a satisfactory description of the linear-response regime if it is to be of use in more general quantum conductance calculations.
Apart from the  
ultra-non-local contribution found by Sai {\it et al.}~\cite{Sai05}, we identify 
a new correction to the conductance that already appears within the adiabatic 
local-density approximation (ALDA), and gives a significant increase in the conductance even for the homogeneous electron gas.
By using an expression for the conductance in real space, we 
can explore both the ALDA correction and the correction of Sai {\it et al.} for homogeneous and inhomogeneous systems, in relation to the gDFT conductance. 

The central concept in our approach~\cite{Bokes04} is the identification 
of the conductance $G$ as the strength of the Drude singularity of 
the conductivity tensor in reciprocal space\footnote{We use atomic 
units where $e=\hbar=m_e=1$.}, 
\begin{equation}  \label{eq-sigma}
	\lim_{\omega \rightarrow 0} \sigma_{zz}(q,q',\mathbf{K}_{\perp}=0;\omega) =
	2\pi G \delta(q) \delta(q'),
\end{equation}
where we consider a geometry where $z$ is the direction of the current flow,
$q$ a reciprocal vector in that direction, and for simplicity we assume
that the system is translationally invariant along the $x,y$ 
directions and $\mathbf{K}_\perp$ is a wavevector reciprocal to 
$\mathbf{R}_\perp = (x,y)$
, i.e.  we consider an ideal interface. The conductance $G$ is then a conductance per unit area of the interface. 
The limit $\omega \rightarrow 0$ must be taken from
the upper half of the complex frequency plane $\omega=\Re\{\omega\} + i\alpha, 
\alpha>0$ as the last step in the calculation. This order of limits -- 
first an infinitely long system (i.e. $q$ continuous), and only then
$\omega \rightarrow 0$ -- is essential, for otherwise different and
even divergent results are obtained, arising from the incorrect 
``piling-up'' of charge at the ends of the system~\cite{Bokes06}. The 
unambiguous evaluation of a finite conductance for an infinite dissipationless 
system is facilitated by the adiabatic switching-on of an external 
electric field, characterized by a finite drop in potential $\Delta V^{ext}$,
that can conveniently~\footnote{We exploit the insensitivity of the the steady-state conductance to an additive $\Delta E^{ext}_{z}(z,t); \int \Delta E^{ext}_{z}(z,t) dz =0$, arising from the singular character of the conductivity.} 
be represented as the field
\begin{eqnarray} 
	E^{ext}_{z}(z,t) = - \frac{\Delta V^{ext}}{\pi} 
		\frac{c / \alpha}{(c / \alpha)^2 + z^2}  e^{-i\omega t},
\end{eqnarray}
for the time interval $t\in(-\infty,0)$. $c$ is a constant with units of velocity 
and $\alpha>0$ controls the speed of switching on the field.  

Through Eq.~(\ref{eq-sigma}) we can relate the
conductance to the non-local conductivity 
for small $\omega$.  In turn, the conductivity is simply related to 
the irreducible polarization~\cite{Bokes04,Pines67}~\footnote{
Henceforth we suppress the explicit appearance of $\mathbf{K}_\perp=0$ in the density responses $\chi$.}
\begin{eqnarray}  \label{eq-sigma-chi}
	\sigma_{zz}(q,q',\mathbf{K}_\perp=0;\omega) = \frac{i\omega}{qq'} \chi^{irr}(q,q';\omega).
\end{eqnarray}
The polarization is conveniently calculated via the density response 
function calculation within TDDFT (although other possibilities exist, such as 
TDCDFT or many-body perturbation theory).
The polarization function satisfies the Dyson equation
\begin{eqnarray} 
	\chi^{irr}(q,q';\omega) = \chi^0(q,q';\omega) + \nonumber \\ 
		\int dq'' dq'''
		\chi^{0}(q,q'';\omega) f_{xc}(q'',q''';\omega) 
		\chi^{irr}(q''',q';\omega) \label{eq-chi-dyson}, \label{eq-4}
\end{eqnarray}
where $\chi^0(q,q';\omega)$ is the non-interacting Kohn-Sham density response function.
The above two equations give the conductance of an interacting electronic system: 
for a given kernel $f_{xc}$ one needs to invert the Dyson equation (\ref{eq-4}) 
and substitute the result into Eq.~(\ref{eq-sigma-chi}). 

However, we can gain insight by multiplying Eq.~(\ref{eq-chi-dyson}) 
by $i\omega/(qq')$ and taking the limit $\omega \rightarrow 0$. Clearly, 
the resulting left-hand-side is singular in $q$, with the strength being directly 
the conductance of the interacting system, $G$. The strength of the first 
term on the right-hand-side of (\ref{eq-4}) multiplied by the same factor 
gives the conductance of the non-interacting Kohn-Sham system, $G^{0}$. The difference 
between these 
two conductances, 
i.e. the correction due to the exchange-correlation kernel,
is then nonzero \emph{only if} the last term in Eq.~(\ref{eq-chi-dyson}) 
also leads to a singular form. This observation can be used to deduce 
the aspects of the kernel that \emph{do} influence the conductance, since 
the general character of $\chi$ for small $\omega$ is well known. 

The most obvious choice, making use of the character of $\chi^{0/ir} \sim qq'$,
is $f_{xc}^{(a)}(q,q';\omega) = -\frac{i\omega}{qq'} A(q,q'; \omega)$ 
where $A(q,q';\omega) \rightarrow A = A(q=0,q'=0) \neq 0$ for $\omega \rightarrow 0$.
The resulting conductance then takes the form
\begin{eqnarray} 
	G = G^0 - G^0 A G = \frac{G^0}{1 + A G^0}. \label{eq-G-A}
\end{eqnarray}
From the above expression we see that $A$ represents some part
of the dynamical resistance per unit area and, in fact, it is equivalent to the
correction identified by Na Sai {\it et al.}~\cite{Sai05}. It can be shown~\cite{Bokes06} 
that a purely longitudinal exchange-correlation electric field used in their 
treatment~\cite{Vignale96,Vignale97} within TDCDFT is equivalent to 
a contribution to the TDDFT kernel of the asymptotic form for small $q,q'$
\begin{eqnarray} 
	f^{(a)}_{xc}(q,q';\omega) \approx 
		- \frac{i\omega}{q q'} \int dz \frac{4\eta}{3} 
		\left( \frac{\partial_z n(z)}{n(z)^2}  \right)^2 
	=	- \frac{i\omega}{q q'} A  \label{eq-A}
\end{eqnarray}
where $n(z)$ is the number of electrons per unit cross-sectional area and 
$\eta$ is the dynamical viscosity of a homogeneous electron 
gas~\cite{Vignale97}. 
We should note that for homogeneous systems $A=0$ since $\partial_z n(z)=0$. 
This is important since the functional form $f_{xc}^{(a)}$ given above and 
the limiting process would not lead to a finite result for a homogeneous system.

However, the form above does not exhaust all the possibilities.
Most surprisingly, the conductance is also affected by a {\it local-density} term  
$f_{xc}^{(b)}(q,q';\omega) = f_{xc}^{(b)}(q-q';\omega) \rightarrow B(\omega) \delta(q-q'), 
B(\omega) \rightarrow B \neq 0$ for $\omega \rightarrow 0$, 
which, Fourier-transforming $f_{xc}^{(b)}$ to real space, 
naturally appears within the adiabatic LDA~\cite{Gross85} in an extended system,
\begin{eqnarray} 
	f_{xc}^{(b)}(\mathbf{r},\mathbf{r}') = f_{xc}^{ALDA}[n^0(\mathbf{r})] 
		\delta (\mathbf{r}-\mathbf{r}') \\
	f_{xc}^{ALDA}[n^0]= d^2 ( n^0 \epsilon_{xc}(n^0) ) / dn^2 .
	\label{eq-8}
\end{eqnarray}
(Here $\epsilon_{xc}(n^0)$ is the exchange-correlation energy per particle of a 
homogeneous electron gas (HEG) of density $n^0$). 
To exhibit this case let us consider the 1D HEG where the algebra becomes 
particularly simple, and we will assume that the kernel 
$B_{1D}=f^{1D}_{xc}[n^0]$ exists even for this non-Fermi-liquid system. 
(For 2D and 3D cases, where this fact is generally accepted, 
the analogous steps lead to technically more demanding formulas and only some 
of the results can be obtained analytically. However, we will demonstrate numerically 
that the qualitative general picture is identical to that described below for 
the 1D situation.)

The non-interacting response of the 1D gas, $\chi^0_{\rm 1D}$, for $q << k_F$, has the simple form
\begin{eqnarray} 
	\chi^0_{\rm 1D}(q,q';\omega) \approx -\frac{2}{\pi k_F} 
		\frac{k_F^2 q^2}{
			k_F^2 q^2 - \omega^2 } \times \delta(q-q'). 
\end{eqnarray}
When $f_{xc}^{(b)}$ is combined with this in the Dyson equation (\ref{eq-4}), 
a renormalization of the conductivity results,
\begin{eqnarray}  \label{eq-sigma-heg}
	\sigma_{\rm 1D}(q,q';\omega) \approx  
	\frac{2 k_F}{\pi} \frac{-i \omega}{
			(1+\frac{2B_{\rm 1D}}{\pi k_F} ) k_F^2 q^2 - \omega^2 }
			 \times \delta(q-q'), \quad
\end{eqnarray}
which, using Eq.~(\ref{eq-sigma}), gives the conductance
\begin{equation} 
	G_{\rm 1D} = \frac{G^0_{\rm 1D}}{\sqrt{1+\frac{2B_{\rm 1D}}{\pi k_F}}} 
\end{equation}
Typically (i.e. in 2D and 3D) the adiabatic kernel (\ref{eq-8}) 
is negative and decreasing function of the density and eventually 
as a certain critical density (characterized by $k_{F,c}$ given by 
$\frac{2}{\pi k_F}B_{\rm 1D}(k_{F,c}) = -1$) is approached, 
the conductance shows a singular increase.  This transition corresponds to
the known instability
of the HEG against arbitrarily small fluctuations in the total 
potential~\cite{Pines67}, signified by the appearance of a pole in the upper half of the complex plane 
of the irreducible density response $\chi^{irr}$.  For densities beyond the transition, linear-response theory
is therefore inapplicable.

\begin{figure}
\includegraphics[width=8cm]{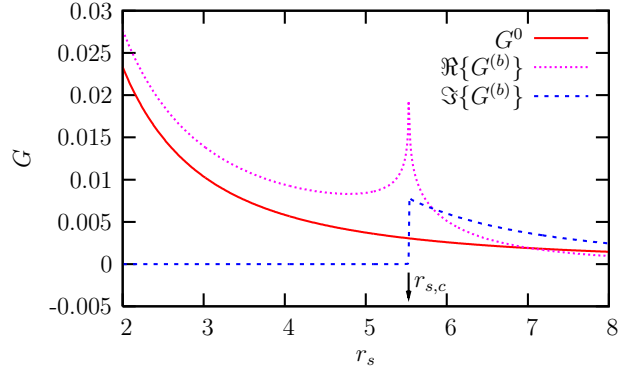}
\caption{(Color-online) Dependence of the conductance per area of the Kohn-Sham gas in 3D, 
$G^0$, and the real and imaginary parts of the full conductance within the ALDA, ($G^{(b)}$), 
on the density parameter, $r_s$. The ALDA correction is significant within the 
physically relevant region $2 \lesssim r_s<r_{s,c}$.} \label{fig-1}
\end{figure}

The situation in 3D gas is qualitatively similar. For $q << q_{F}$ 
the non-interacting response has the form ($\omega=i\alpha, \alpha>0$)
\begin{equation} 
	\chi^0(q,i\alpha) = - \frac{k_F}{\pi^2} \left( 
	1 - \frac{\alpha}{q k_F} \arctan \frac{k_F q}{\alpha} \right),
\end{equation}
from which, by means of Eqs.~(\ref{eq-sigma}) and (\ref{eq-4}), we obtain the 
conductance per unit area of the Kohn-Sham gas $G^0=\frac{k_F^2}{4\pi^2}$. 
The critical density at which the instability occurs can be found 
from the appearance of poles of the the irreducible 
response function in the upper half of the complex plane, which leads to the 
criterion $B(k_{F,c}) \leq - \frac{\pi^2}{k_{F,c}}$. The form of 
the correction to the conductance in 3D cannot be obtained analytically, 
so we show in Figure~\ref{fig-1} our numerical results, which qualitatively resemble 
the 1D case. We see that 
the ALDA correction leads to a systematic increase in the conductance. 
We stress that this correction is a direct consequence of the 
proper order of limits performed in Eq.~\ref{eq-4}.  From this it also 
follows that this correction would \textit{not} be present within the static gDFT 
calculations based on the NEGF formulation but \textit{would} appear within a direct 
time-evolution approach~\cite{Kurth05,Bushong05} if the ALDA is employed.
(A minor difference in 3D is that beyond the critical $r_{s,c}$ 
the formally defined conductance retains 
a nonzero real part, whereas in 1D it is pure imaginary.)

We briefly discuss a third form of the exchange-correlation kernel, 
$f_{xc}^{(c)}(q,q';\omega) = \frac{\omega^2}{q^2} C(q,\omega)
\delta(q-q'), \quad C(q,\omega) \rightarrow C \neq 0$ 
for $\omega,q \rightarrow 0$. It can be shown that this form of frequency dependence in the kernel also changes the final conductance. 
This kernel is similar to the one used by Botti {\it et al.}~\cite{Botti05} 
and Reining {\it et al.}~\cite{Reining02}
for bulk insulators and semiconductors characterized by a finite gap in 
the electronic density of states at the Fermi energy. In their estimate, 
which lead to significant improvement in the electron energy-loss spectra 
and optical absorption spectra, the coefficient $C \sim \omega_g$ 
where $\omega_g$ is the effective energy gap. 
This suggests that $f_{xc}^{(c)} = 0$ will be zero, or small, in a metallic system. 

In order to explore the relevance of these conductance corrections for inhomogeneous systems
we consider the metal-vacuum-metal case of  
two jellium surfaces separated by a distance $d$. In this case the parameter 
$B$ has been evaluated using Eq.~(\ref{eq-8}) and the 
Perdew-Zunger~\cite{Perdew81} parametrization of the quantum Monte Carlo 
correlation energy of the HEG~\cite{Ceperley80}. The parameter 
$A$ is obtained from Eq.~(\ref{eq-A}), and its contribution to the conductance follows from Eq. (\ref{eq-G-A}). 

\begin{figure}
\includegraphics[width=8cm]{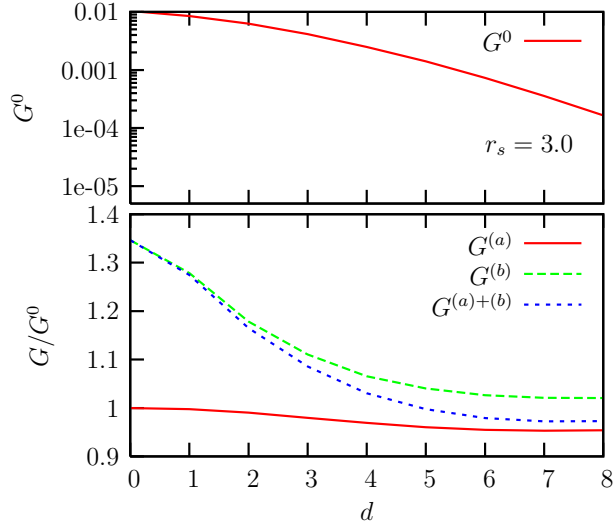} 
\caption{(Color-online) 
The Kohn-Sham conductance $G^0$ (upper panel)
and the corrected conductances $G$ (shown relative to $G^0$) (lower panel), as a function of the 
vacuum width $d$. The corrections due to the non-local (a) and ALDA (b) kernels 
have opposing signs.}
\label{fig-4}
\end{figure}

In our calculations we employ two jellium 
slabs of thickness $L$ and density given by $r_s$.
The calculation of $\chi^0(z,z';i\alpha)$ is performed at the self-consistent LDA level 
~\cite{Jung04}. 
Subsequently we invert the Dyson equation, Eq.~(\ref{eq-4}), 
in real space and thereby calculate the irreducible response 
$\chi^{irr}(z,z';i\alpha)$.
For the final step we employ an integral form of Eq.~(\ref{eq-sigma})
\begin{equation} 
	G = - \lim_{\alpha \rightarrow 0^+} \alpha \int \frac{dq dq'}{2\pi} 
		\frac{\chi^{irr}(q,q';i\alpha)}{q q'}, 
		\label{eq-G-qq}
\end{equation}
where we have used the identity (\ref{eq-sigma-chi}). Direct implementation of this expression
in reciprocal space is numerically rather cumbersome, since
$\chi^{irr}$ is, in practice, calculated for a finite (periodic) supercell for which 
the region with $q,q'<<1$ is poorly described. This problem
can be circumvented by re-expressing $G$ in a real-space form,
\begin{eqnarray} 
	G = - \lim_{\alpha \rightarrow 0^+} \alpha 
		\int_0^{\infty} dz \int_{-\infty}^{0} dz'
		\chi^{irr}(z,z';i\alpha), \label{eq-G-zz}
\end{eqnarray} 
which is obtained from (\ref{eq-G-qq}) by utilizing the fact that 
the $1/qq'$ singularity is only apparent since $\chi(q,q') \sim qq'$ 
for small $q,q'$. Extrapolation to zero frequency 
(from  $\omega \sim 0.01$ Ha $<< E_F$) is done in parallel with 
extrapolation of 
the thickness of the slabs $L$ to infinity. Further details of the calculation 
will be published elsewhere~\cite{Bokes06}. The quality of the numerical procedure can be judged from the correct exponential decrease in the 
Kohn-Sham conductance over several orders of magnitude shown in 
Fig.~\ref{fig-4}.

The resulting dependence of the conductance on the vacuum width 
is shown in Fig.~\ref{fig-4}, for a representative jellium density 
$r_s=3$~a.u.~(Au).
For small vacuum widths $d$, the correction due to the ALDA kernel 
$f_{xc}^{(b)}$ clearly dominates;
for larger widths ($d \sim 6 - 8$ a.u.) $f_{xc}^{(a)}$ and $f_{xc}^{(b)}$ 
shift the Kohn-Sham conductance in opposite directions and to some extent 
cancel each other. We should note that unlike the use of a global value 
for the viscosity in the original work by Sai~\cite{Sai05}, we have used 
the local viscosity $\eta[k_F(z)]$ in Eq.~(\ref{eq-A}), determined by 
the local density, which leads to suppression of the non-local corrections 
for $d \gtrsim 2$a.u.\cite{Bokes06} 

In conclusion, we have presented a unified formalism, based on the singular character of response 
functions, to address the 
conductance of a general system of interacting electrons. We have explicitly 
identified three different contributions to the dynamical resistance:
(a) the non-local contribution~\cite{Sai05} parametrized by the dynamical viscosity 
of the homogeneous electron gas, effective only for inhomogeneous systems,
(b) a local contribution parametrized by the adiabatic LDA 
exchange-correlation kernel, and (c) an ultra-non-local contribution that, 
from presently available estimates, is not important for metallic systems. 
These three forms are unlikely to exhaust all the possibilities, 
and our theoretical framework remains applicable for further analysis based on many-body perturbation theory to obtain 
additional relevant contributions to the exchange-correlation kernel or local field factor.
In the homogeneous electron gas, we have found that the conductance diverges as 
the critical density $r_{s,c}$ is approached, beyond which the gas is unstable.
For practical calculations, our formalism can readily benefit from available 
computer codes for \textit{ab-initio} calculations in real materials that can generate 
density-density response functions.  We have shown the importance of the non-vanishing 
corrections for the conductance of a model inhomogeneous system.

\begin{acknowledgments}
This research was supported by a Marie Curie European Reintegration Grant
within the 6th European Community RTD Framework Programme (QuaTraFo,
MERG-CT-2004-510615), the Slovak grant agency VEGA (project No. 1/2020/05),
the NATO Security Through Science Programme (EAP.RIG.981521) and the EU's 6th
Framework Programme through the NANOQUANTA Network of Excellence
(NMP4-CT-2004-500198).
\end{acknowledgments}


\end{document}